\def\year{2019}\relax
\documentclass[letterpaper]{article} 
\usepackage{aaai19}  
\usepackage{times}  
\usepackage{helvet}  
\usepackage{courier}  
\usepackage{url}  
\usepackage{graphicx}  
\usepackage[utf8]{inputenc}

\usepackage{multirow}
\usepackage{hhline}
\usepackage{multicol}
\usepackage{graphicx}
\usepackage{epsfig}
\usepackage{multirow}
\usepackage{amssymb}
\usepackage{amsmath}
\usepackage{dsfont}
\usepackage{bbm}
\usepackage[ruled,linesnumbered]{algorithm2e}

\usepackage{mathtools}

\usepackage[left]{lineno}
\usepackage{blindtext}

\usepackage{ upgreek }
\usepackage{caption}
\usepackage{subfig}

\usepackage{natbib}
\usepackage{graphicx}
\usepackage{booktabs} 

\copyrightyear{2019}

\begin{document}
\title{Modeling the Dynamics of User Preferences for Sequence-Aware Recommendation Using Hidden Markov Models}


\author{Farzad Eskandanian \\
feskanda@depaul.edu \\
Center for Web Intelligence \\ DePaul University \\
Chicago, IL, USA \\
\And Bamshad Mobasher \\
mobasher@cs.depaul.edu \\
Center for Web Intelligence \\ DePaul University \\
Chicago, IL, USA}

\maketitle

\begin{abstract}
In a variety of online settings involving interaction with end-users it is critical for the systems to adapt to changes in user preferences. User preferences on items tend to change over time due to a variety of factors such as change in context, the task being performed, or other short-term or long-term external factors. Recommender systems need to be able to capture these dynamics in user preferences in order to remain tuned to the most current interests of users. In this work we present a recommendation framework which takes into account the dynamics of user preferences. We propose an approach based on Hidden Markov Models (HMM) to identify change-points in the sequence of user interactions which reflect significant changes in preference according to the sequential behavior of all the users in the data. The proposed framework leverages the identified change points to generate recommendations using a sequence-aware non-negative matrix factorization model. We empirically demonstrate the effectiveness of the HMM-based change detection method as compared to standard baseline methods. Additionally, we evaluate the performance of the proposed recommendation method and show that it compares favorably to state-of-the-art sequence-aware recommendation models. 

\end{abstract}

\section{Introduction}

Traditional methods for personalized recommendation involve learning from long-term preferences of users and tailoring the recommendations to new users based on their overall preference profiles. This approach, however, does not take into account the fact that a user's preferences change over time and items that may have been relevant or of interest in the past may no longer suit the needs of the user in the present. Changes in a user's preferences may occur because of a variety of factors including changes in user's situation, context, the task at hand, or even due to one-time external events.

This problem is particularly pronounced in domains where a user's interest in items may vary often in the course of interactions with the system. An example of such a domain is music streaming where a user's sequence of interactions with the system (such as selecting, skipping, liking or disliking a song) is recorded and some or all of this history of interactions is used to identify future items to present to the user. But, even in domains with less transient behavioral characteristics, user preferences may change over time because a user's tastes may evolve slowly and thus older items in the user profile may no longer reflect the user's current preferences \citep{mcauley2013amateurs}. Examples of such a situation may be user preferences for wine where a user's taste may evolve over time due to experience and the development of a more discerning palate. 

To address this problem, recommender systems must be able to model the dynamics of user preferences by identifying change points in the user interaction sequence beyond which a user's behavior might indicate a significant change in preferences, and finally to tailor the system's recommendations to the most relevant episodes within the user's overall profile associated with the identified change points.  

Recent research in recommender systems has tried to address different aspects of this problem. For example, context-aware recommender systems (CARS) \citep{adomaviciusCARS2011, haririMusicContext2012, karatzoglou2010multiverse, zheng2014cslim} try to take into account the current context of the user or the most appropriate context for an item when generating recommendations. Most common approaches to CARS, however, rely on explicit representation of contextual factors which are not always available and generally do not try to model the dynamics of user preferences. Furthermore, session-based recommender systems \citep{jannach2017recurrent, hidasi2015session} have been introduced with a focus on developing models that generate recommendations using only the observed behavior of a user during an ongoing session while ignoring parts of user's overall profile that are considered to be associated with previous sessions. While session-based recommenders address part of the aforementioned problem by trying to model sequential interactions with the system, they generally do not address the problem of explicitly modeling and automatically detecting change points in user preference sequences. 




In this work we present a recommendation framework based on Hidden Markov Models (HMM) that integrates automatic change point detection within sequences of user interactions with recommendation models that take into account the dynamics of user preferences. We specifically focus on a setting where user preferences are implicitly and sequentially captured during the course of a user's interaction with the system and where there is no explicit representation for contextual or other factors that may provide {\em a priori} indications of possible changes in user preferences. Our goal is to automatically identify change points in user preferences and use them to generate sequential recommendation. We conjecture that appropriate identification of change points and tailoring recommendations to the most relevant segments in the history of user interactions will lead to more effective recommendation. 

Given a sequence of user-item interactions, an HMM can be used to identify the most likely sequences of hidden states representing change points in user preferences. This change point detection mechanism can be used to identify specific segments of user-item interaction sequences to be used in as input for a traditional non-sequential recommendation model such as matrix factorization. This approach, in and of itself, should lead to more effective recommendation when compared to the same approach without change point detection. However, the advantage of using HMMs is that the same learned model can also be used to infer probabilities associated with items in the observation sequences which in turn can be used to directly generate recommendations for next items without resorting to other non-sequential recommendation models \citep{eskandanian2018detecting}.

The center piece of our proposed framework is the change point detection mechanism using an HMM. The identified change points are used to segment user sequences with different segments representing significantly different preference models. We then integrate this change detection method into a standard recommendation model using non-negative matrix factorization. 

We empirically evaluate our approach using two real music streaming data sets from Spotify, Inc. and Lastfm. We show that this principled approach for modeling preference dynamics leads to more effective recommendations than other baseline approaches to sequence-aware or session-based recommendation (such as those using recurrent neural networks).

\section{Related Work}

There are three general areas of research in recommender systems that address the problem considering changes in user preferences, interests, or situations. One is Context-Aware Recommendation Systems (CARS) \citep{hariri2013query, adomaviciusCARS2011}. Most CARS approaches assume a predefined representational view to modeling context and the main goal is to leverage contextual information in generating recommendations. 

In other work, the case where contextual information is not observable directly and has to be inferred has been studied \citep{haririMusicContext2012}. However, these approaches are devised to infer the contextual information implicitly using topic modeling of content information such as tags associated with items. Also, in \citep{TompsonSampling_hariri2014context}, the authors have proposed to use a multi-armed bandit algorithm in order to quickly adapt the recommendations to the contextual changes in the case of interactive recommendations in an online setting. The main focus of that work is to apply bandit algorithms in interactive recommender systems and the change point detection is essentially an add-on feature independent of the learned model. 

In other work, Markov Models have been used to model users' sequential behavior  \citep{HHMM2015, rendleFPMC2010, FOSSIL2016, rendleFPMC2010, abdollahpouri2017towards}. In \citep{HHMM2015} a Hierarchical Hidden Markov Model is used to implicitly model the hidden states as context and generate recommendations using the last inferred context of user. Although, same approaches are used to model the dynamics of user preference, our model is based on single hidden layer compared to a Hierarchical hidden structure. Besides simplicity, our work extends Viterbi's algorithm for decoding task and generates recommendations using latent factors learned from Emission probabilities. The simplicity of our model compared to \citep{HHMM2015} makes it more efficient in training time. There are also other recent approaches \citep{FOSSIL2016, rendleFPMC2010} that are similar to our approach in terms of their overall goal of modeling user dynamics, but different in terms of their methodology.

\section{Problem Definition and Background}

In many recommendation domains users interact with a collection of available items through various actions such as viewing, clicking, or selecting items. These user-item interaction sequences form the underlying observations in our HMM-based change detection and recommendation framework.  

Let $\mathcal{U} = \{u_1, u_2, ..., u_N\}$ be a set of users and $\mathcal{I}$ = $\{i_1, i_2,$ $...$ $i_M\}$ set of items. For each user $u$ the list of his/her interactions is denoted by $I_u$ = $\langle i^{(1)},$ $i^{(2)}$ $,..., $ $i^{(T)}\rangle$. Each element $i^{(t)}$ in this list represents the interaction of user $u$ with item $i$ at sequential index $t$ and all of the items are ordered based on the time of interaction. 

A change point $\Lambda_j$ in the sequence of user interactions is an index that partitions this sequence into $I_1 = \langle i^{(1)},... i^{(\Lambda_j)} \rangle$ and $I_2 = \langle i^{(\Lambda_j+1)},...i^{(T)} \rangle$. In general, there is no limit for the number of change points and the resulting segments and there could be multiple change points $\Lambda = \{ \Lambda_1, ..., \Lambda_k\}$ within one user sequence depending on the significance of changes in preferences. The challenge here is to accurately detect the set of change points $\Lambda$. More precisely, $\Lambda$ should partition the sequence of user interaction in a way that maximizes intra-partition similarity and minimizes inter-partition similarity of $\langle I_0$ $,...,$ $I_k\rangle$. 


\subsection{Hidden Markov Models}

Hidden Markov Models (HMM) belong to the category of probabilistic models specifically used for modeling sequential data. Markov Chains (MC) are the simplest in this category which are based on following independence assumption known as Markov property. Given a set of discrete time-based variables $Y \in \{ y_1$, $y_2$, ...$y_n\}$, and a sequence of these variables $\langle Y_1$, ..., $Y_{t-1}$, $Y_t \rangle$, in a first-order Markov model the probability of the $Y_t$ after seeing the sequence depends only on the last observation $P(Y_t = y|Y_1$, ..., $Y_{t-2}$, $Y_{t-1})$ = $P(Y_{t} = y|Y_{t-1})$. In recommendation domain due to the large number of items and hence the resulting data sparsity, Markov Chain models tend to perform poorly. Higher-order Markov Chains can be used, but at the cost of significantly higher time and space complexity.

Hidden Markov Models are extensions of Markov Chains that model the sequential patterns in the data using the transition probabilities between the hidden states instead of observations. The underlying assumption is that the transition probabilities among hidden states $Z \in \{z_1$, ..., $z_k\}$ cannot be inferred directly using observations. Formally, an HMM model is defined by an initial hidden state distribution $\pi$, state transition probabilities $\mathcal{A} = P(Z_t|Z_{t-1})$, and emission probabilities $\mathcal{B} = P(Y_t|Z_t)$. We denote an HMM by $\theta = (\mathcal{A}, \mathcal{B}, \pi)$.   

HMMs have been used to solve three different general problems that make these models very practical and suitable in many situations \citep{gales2008application}.
\begin{enumerate}
\item \textbf{Likelihood estimation problem}: Given a sequence of observations $\langle Y_1$, ..., $Y_{t-1}$, $Y_t \rangle$ , and $\theta$, determine the $P(Y_1$, ...$Y_t|\theta)$.
\item \textbf{Decoding problem}: Given a sequence of observations and $\theta$, what is the most likely sequence of hidden states: $\operatorname*{arg\,max}_{Z_1,...Z_t}{P(Z_1,...,Z_t,Y_1,...,Y_t|\theta)}$
\item \textbf{Learning problem}: Given a sequence of observations and a set of hidden states, learn HMM parameters $\theta$ using maximum likelihood estimation.
\end{enumerate}

In the following section we demonstrate how we can adapt these problems and their solutions to the problem of detecting changes in user preferences over time. The identified change points can then be used in the recommendation phase taking into account shifting user preferences.

\section{Proposed Framework}

\subsection{Change Point Detection}

Given a sequence of user-item interactions (implicitly representing user's preferences on items), we first learn $\theta = (\mathcal{A}, \mathcal{B}, \pi)$. We use the well-known \textit{Baum-Welch} algorithm to learn the model from the data \citep{BaumWelch1986}. This method is based on the Expectation-Maximization (EM) algorithm to find the maximum likelihood estimate of $\theta$ using the sequence of observations. Next, using the learned model we ``decode" the hidden states related to each observation (i.e., each item in the interaction sequence). The standard algorithm for this task is the \textit{Viterbi} algorithm \citep{viterbi}. Given a sequence of observations and $\theta$, \textit{Viterbi} algorithm uses a dynamic programming approach to find the the most likely hidden states corresponding to the sequence of observations. It is this generated sequence of hidden states which represents the dynamics of user preferences over time. 

In order to identify a change point $\Lambda$ we use the \textit{Viterbi} algorithm as follows. \textit{Viterbi} relies on $\theta$ to find the maximum path $\mathcal{V} = \langle v_1,...,v_T\rangle$ among the hidden states corresponding to each item in the sequence of observations. Suppose that the hidden state corresponding to the \textit{Viterbi}'s path at time $t$ is denoted by $\mathcal{V}_t$. A change is detected if there is a hidden state $\mathcal{V}_t$ with maximum tendency to change from $\mathcal{V}_{t-1}$, that is: $\operatorname*{arg\,max}_{t}{P(\mathcal{V}_t|\mathcal{V}_{t-1}) P(i_t|\mathcal{V}_t)}$ Where $\mathcal{V}_t \neq \mathcal{V}_{t-1}$.
 
However, this approach only detects one change point which has the highest probability. Also, the change point in Viterbi's path with maximum probability does not necessarily correspond to a significant change in preference. 
In order to address this issues, we add a threshold parameter $\tau$ to the change detection equation:
\begin{equation} \label{eq:MultChangeDetection}
    \Lambda' = \{ t | (\mathcal{V}_t \neq \mathcal{V}_{t-1}) \hspace{0.2em} \land \hspace{0.2em} P(\mathcal{V}_t|\mathcal{V}_{t-1}) P(i^t|\mathcal{V}_t) > \tau \}
\end{equation}

Since the end goal of detecting the changes in the user preference is to increase the effectiveness of recommendations and hence is increasing accuracy, we tune the parameter $\tau$ so that the maximum accuracy can be achieved. 


\begin{algorithm}[t]
\caption{Hidden Markov Change point Detection HMCD}
\label{alg_hmm_change_detection}
\KwIn{Set of sequences of User-Item interactions $X = \{ I_{u_1},...,I_{u_n} \}$; Threshold for change detection $\tau$; Number of Hidden States $h$.}
\KwOut{Partitioned User-Item Matrix $M$.}
    {$M$: User-item interaction matrix $M \in \{0, 1\}^{n'\times m}$, where $n' = n \times k$.}
    
    \tcc{Learn HMM Model.}
    {$\theta \gets \textit{Baum-Welch}(X, \hspace{0.2em} h)$}
    
    \For {$u \in \mathcal{U}$} {
    \tcc{Using \textit{Viterbi}, Decode each item $i \in I_u$.}
    {$\mathcal{V} \gets$ \hspace{0.2em} $Viterbi(\theta, \hspace{0.2em} I_u)$}
    
    \vspace{0.5em}
    
    \tcc{Find time index of State Changes in $\mathcal{V}$'s path that pass the probability threshold $\tau$.}
    $\Lambda' = \{ t | (\mathcal{V}_t \neq \mathcal{V}_{t-1}) \hspace{0.2em} \land \hspace{0.2em} P(\mathcal{V}_t|\mathcal{V}_{t-1}) P(i^t|\mathcal{V}_t) > \tau \}$
    
    \vspace{0.5em}
    \tcc{Partition $I_u$ into $k+1$ parts where $k=|\Lambda'|$.}
    {$\langle I^0_u,...,I^k_u \rangle = Partition(I_u, \hspace{0.2em} \Lambda')$}
    
    \For {$i=0$ \KwTo $k$}{
        $M_{u\times k+i} = I^i_u$
    }
    }
{return $M$}
\end{algorithm}

\subsection{Generating Recommendations Using Detected Change Points} \label{recommendation_approach}

We call our approach for generating recommendations \textbf{Sequence-based Matrix Factorization (SMF)}. In this method a Matrix Factorization model is used based on detected change points in the sequence of user interactions. After decoding the hidden states of each item in $I_u$ and segmenting this sequence using Algorithm~ \ref{alg_hmm_change_detection}, we treat each segment of items as a user profile vector used in the factorization model. More formally, given $I_u$ the sequence of interactions of user $u$, the HMM decoding is used to identify a list of change points $\Lambda'$. Then, $I_u$ will be segmented into $\langle i^{(1)},... i^{(\Lambda'_1)} \rangle$ $, ...,$ $\langle i^{(\Lambda'_k+1)},...i^{(T)} \rangle$. Where the segments are denoted by $\langle I_u^0, ..., I_u^k\rangle$, and the list of $k$ predicted change points is denoted by $\Lambda'$. The new segments for every user are used to generate a new user-item interaction matrix $M$. To find approximate factorization of $M \approx p.q^T$, we use Non-negative matrix factorization. The objective in this method is to minimize the Euclidean distance between the approximated matrix $p.q^T$ and the actual matrix $M$. More precisely, NMF minimizes $|| M - p.q^T ||^2$ with respect to $p$ and $q$, subject to $p,q \geqslant 0$. Since, change point detection has decoupled the items of each user profile in $M$ which sequentially do not belong to the same hidden state, the association of items with sequential likelihood in matrix factorization will not be lost. Therefore, using NMF we can estimate $q^T$, the item factors of $M$, which are based on the sequential patterns of items in the original user-item matrix $X$. Note that as long as we keep the number of change points small, the sparsity as a result of segmentation should not decrease accuracy of recommendations compared to static matrix factorization. 

In order to generate recommendations for $u$, the last segment of the user interactions $I^k_u$, which represents the latest preference of the user will be used (in future work we will explore using other relevant or similar segments in addition to the last one in the user sequence). After training the matrix factorization model the latent factor vector corresponding to the latest interests of user $I^k_u$ will be used to find the items with the best match to this representation: $R_u = \operatorname*{arg\,max}_{j}^{l}{(p_u .q^T_{j})},$ where $p_u$ corresponds to $I^k_u$ which is the latent factor of the latest segment of user $u$'s interaction sequence. The above equation finds the top-$l$ most similar items to $p_u$ using the dot product as a similarity function.

\section{Experiments}

In our experiments, we first evaluate the effectiveness of the HMM-based change point detection when compared to baseline CPD methods. Next we measure the impact of change point detection on recommendation effectiveness. 
We use two HMM models with various number of hidden states labeled as $S2$ and $S10$ which indicate HMM models with $S=2$ and $S=10$ hidden states, respectively. The second set of experiments are designed to evaluate the proposed Sequence-based Matrix Factorization (SMF) recommendation model (presented in previous section). 

\subsection{Datasets}
We used the Spotify Playlist dataset from RecSys Challenge 2018 \footnote{https://recsys-challenge.spotify.com/} for our experiments. We randomly sampled 6,000 playlists from 1 million playlists available in this dataset which have been created by Spotify's users. Our sample contains 6,905 unique Albums, 475,838 playlist-album pairs with 98.9\% sparsity. Also, the average length of playlists is $\approx80$. We chose this dataset because average length playlists are usually focused on one or two moods or genres. Therefore, this characteristic of playlists make them a good candidate for concatenating multiple playlists as a way to simulate change points in the data. The procedure for generating the data is as follows. First, we randomly sampled two playlists $p_1$ and $p_2$ from the pool of 1M playlists. Then, we combined them by selecting a random size window of each playlist. The ground truth for change point is specified by the point where $p_1$ and $p_2$ are concatenated. We use all of the users to measure the accuracy of change point detection task. For the recommendation, we hold the last 10 items of the playlist $p_2$ (in the same order) as testing data for evaluation.    

The second dataset which was scraped from LastFM's website\footnote{https://www.last.fm/}, contains information about 1,000 users. These users on average listen to 7,000 albums in the course of their interactions with LastFM. There are 5,000 items in this dataset which are albums and the dataset is 99.94\% sparse. Unlike the previous dataset where we concatenated different playlists to simulate change points, in this dataset there is no ground truth or predefined change points in user interactions. The main reason to use this dataset is to generate recommendations using the predicted change points and to evaluate recommendation accuracy. The last 40 items in the sequence of interactions from test users are held for evaluating recommendation effectiveness.  




\begin{figure}[tbp]
    \centering
    \includegraphics[scale=0.30]{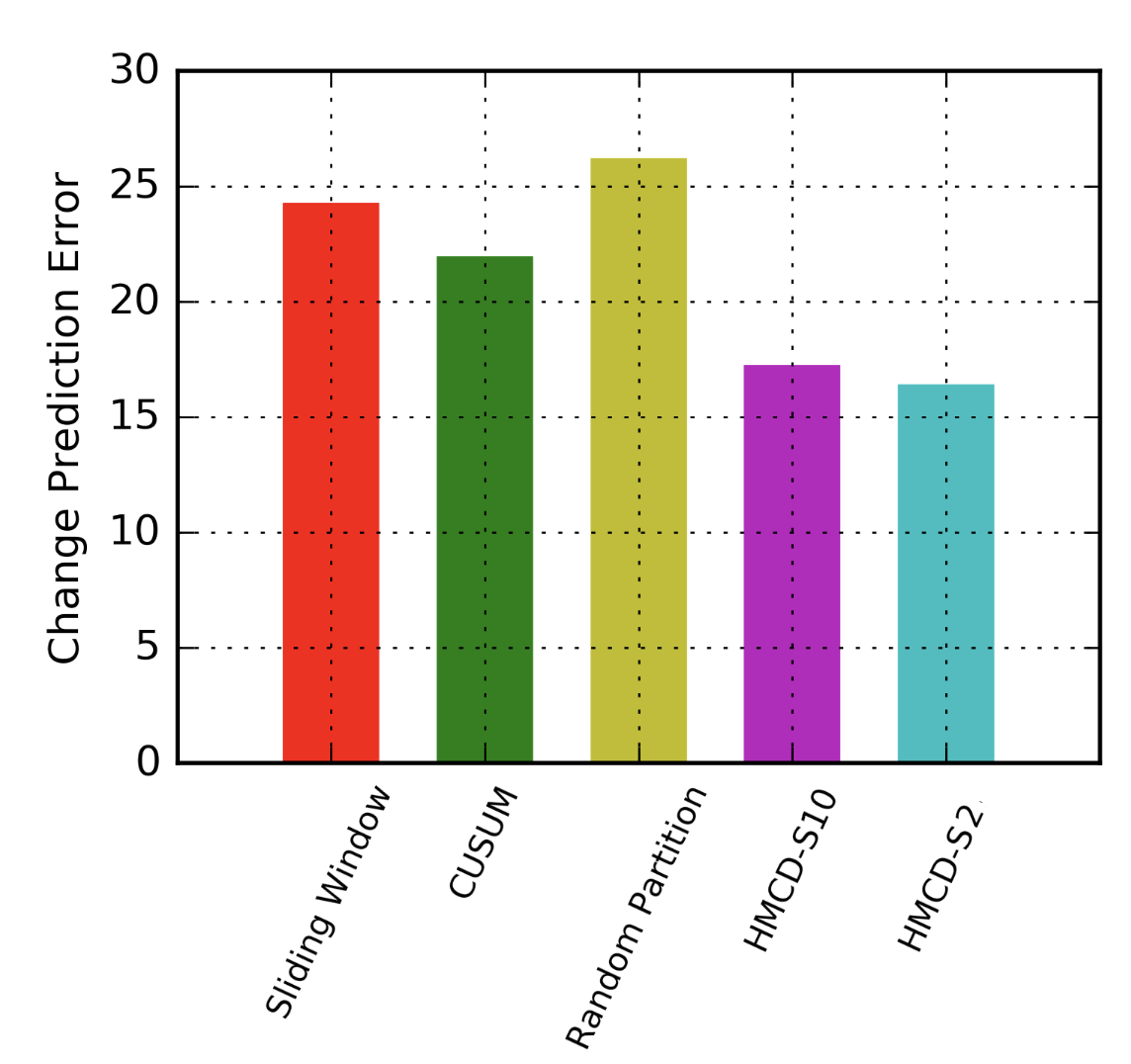}
    \caption{Error of predicted change points using baseline methods and HMMs with different number of hidden states.}
    \label{fig: CH-Acc}
\end{figure}


\begin{table*}[t]
  \centering
  \begin{tabular}{c c c c c c c c c c}
    \hline \hline \\ [-1.5ex]
    Dataset& Method& $P@1$ & $P@5$ & $P@10$ & $R@1$ & $R@5$ & $R@10$ & $nDCG@5$ & $nDCG@10$\\ [0.2ex]
    \hline \\ [-1.9ex]
    \multirow{7}{*}{Spotify Playlist}& 
    PopRank & 4.06 & 3.70 & 2.89 & 0.20 & 0.92 & 1.44 & 0.95 & 1.19 \\
    & MC & 9.67 & 9.91 & \textbf{9.49} & 0.48 & 2.17 & 3.66 & \textbf{8.37} & \textbf{10.47} \\
    & BPR & 8.97 & 6.51 & 6.21 & 0.44 & 1.62 & 3.11 & 1.85 & 2.69 \\
    & GRU & 7.32 & 5.85 & 3.90 & 0.37 & 1.46 & 1.95 & 1.30 & 1.32 \\
    & FOSSIL & 4.02 & 2.97 & 2.57 & 0.20 & 0.74 & 1.49 & 0.77 & 1.01  \\
    & NMF & 11.92 & 9.20 & 7.92 & 0.60 & 2.30 & 3.96 & 3.61 & 4.68 \\
    & SMF-S2($\tau=0.93$) & \textbf{13.46} & \textbf{10.01} & 8.40 & \textbf{0.67} & \textbf{2.50} & \textbf{4.20} & 3.76 & 4.71  \\
    \hline \\ [-1.9ex]
    \multirow{7}{*}{LastFM Listen}&
    PopRank&0.42 & 0.56 & 0.35 & 0.021 & 0.140 & 0.175 & 0.09 & 0.094 \\
    & MC & 9.77 & 9.41 & 8.74 & 0.263 & 1.245 & 2.316 & 2.43 & 3.32 \\
    & BPR & 15.79 & 5.26 & 4.73 & 0.394 & 0.658 & 1.217 & 2.44 & 2.53 \\
    & GRU & 5.22 & 4.42 & 3.82 & 0.26 & 1.112 & 1.915 & 1.54 & 1.98 \\
    & FOSSIL& 12.87 & 12.53 & 10.90 & 0.381 & 1.775 & 3.391 & 2.99 & 4.13 \\
    & NMF& 24.74 & 20.48 & 17.94 & 0.651 & 2.712 & 4.730 & 5.03 & 6.25 \\
    & SMF-S3($\tau=0.9$)  & \textbf{28.29} & \textbf{21.25} & \textbf{18.65} & \textbf{0.759} & \textbf{2.804} & \textbf{4.922} & \textbf{5.53} & \textbf{7.07} \\
    \hline 
  \end{tabular}
  \caption{Percentage accuracy of recommendations at various list sizes (top-1, top-5, and top-10).}
  \label{tab:rec_results}
\end{table*}

\subsection{Change Point Detection Baselines}
We compared the HMM-based change point detection algorithm to several standard change point detection methods often used in time series analysis.

\noindent\textbf{Cumulative Sum, CUSUM}: A well-known approach for change detection in time series. For each item $i \in I_u$ = $\langle i^{(1)},$ $i^{(2)}$ $,..., $ $i^{(T)}\rangle$, CUSUM computes the $S_j = \sum_{t=0}^{j-1}{D(i^{(t)}, i^{(t+1)})}$, where $D(i_1, i_2)$ is the Euclidean distance between two items that are represented using the latent factors (10 factors in our experiments) in NMF. Whenever the cumulative sum exceeds a threshold value $S_j > \tau; \forall  j \leqslant T$,  $\Lambda' = j$ is selected as the change point in the $I_u$. In order to tune the threshold parameter $\tau$, we started with average value of CUSUM among all item sequences and empirically tune  this parameter where the minimum $\Delta$ has been reached.

\noindent\textbf{Sliding Window, SW}:
The sliding partition point (change point) starts from the beginning of the user sequence $I_u$ until it reaches to a point $\Lambda'$ which maximizes intra-partition similarity and minimizes inter-partition similarity. Although, this is a greedy approach, in practice it produces fairly accurate results. We use Euclidean distance as a measure of dissimilarity. 

\noindent\textbf{Random Partition, RP}:
Randomly assign a value to $\Lambda'$ such that $0 \leqslant \Lambda' \leqslant |I^l_u|$.  
\subsection{Recommendation Baselines}

We used several non-sequential as well as recently proposed sequence-aware recommendation approaches as baseline recommendation methods to evaluate the effectiveness of our proposed HMM-based algorithm. In all of the matrix factorization approaches we set the number of latent factors to 40 which provided best results. The baseline methods and some of the parameters used in our experiments are listed below.

\noindent\textbf{FOSSIL}, \citep{FOSSIL2016}: A method  that
fuses similarity-based models with Markov Chains to predict
personalized sequential behavior. Parameters: \texttt{learning rate=0.01; Factors=40;} \texttt{$\alpha$=0.2;} \texttt{regularization=0.} 

\noindent\textbf{GRU}, \citep{hidasi2015session}: A Recurrent Neural Network for session-based recommendation that is widely used as a baseline for sequential recommendation tasks. Parameters: \texttt{loss=top1;} \texttt{number of layers=100-50;} \texttt{update mechanism=Adam;} \texttt{learning rate=0.001}

\noindent\textbf{BPR}, \citep{bprrendle2009}: Bayesian Personalized Ranking optimizes a pairwise ranking objective function via stochastic gradient descent. Parameters: \texttt{learning rate=0.01;} \texttt{Factors=40;}  \texttt{regularization=0.}

\noindent\textbf{NMF}, \citep{NMFlee2001algorithms}: Non-negative Matrix Factorization is another baseline method that decomposes the multivariate data for generating recommendations. NMF uses multiplicative algorithm in order to minimize the least squared error of predictions. Parameters: \texttt{learning rate=0.01;} \texttt{Factors=40;} \texttt{regularization=0.}

\noindent\textbf{MC}: First-order Markov Chain is an item recommendation method based on the sequential patterns captured by transitions from items in all the user interactions.

\noindent\textbf{PopRank}: This method generates a recommendation list for all users by ranking the items based on their popularity among users. Popularity of items in the data are defined by their frequency of being seen/rated by users.

\subsection{Evaluation Metrics}

The error of a single change point prediction is defined by $\Delta = |\Lambda - \Lambda'|$ where $\Lambda$ is the ground truth change point in user sequence of interactions and $\Lambda'$ is the prediction.  As noted before, in our experiments we concatenated different playlists to simulate a change point. Thus the end of one playlist and the beginning of another represent the ground truth change points in our test data. We then computed the average error across all test cases.

In order to measure the ranking accuracy of recommendations, we used Precision, Recall, and normalized Discounted Cumulative Gain (nDCG). The nDCG is time-aware since the gain for items selected earlier by a user are larger when compared to later items. In other words, the true ranking of items are based on the time-based non-decreasing order of items in a user profile. In Table~\ref{tab:rec_results} the accuracy measures are based on various cutoff points in the recommendation lists. For example $P@1$, $R@5$, and $nDCG@10$ are representing Precision, Recall and nDCG of the top-$1$, top-$5$ and top-$10$ items in recommendation lists, respectively.



\subsection{Experimental Results}

The first task in our experiments was to evaluate the accuracy of change point detection. Figure~\ref{fig: CH-Acc} shows the displacement error of change point detection using the baseline methods versus HMCD. We trained two HMM models with 2 and 10 hidden states (HMCD-S2 and HMCD-S10, respectively). Since on average the length of each mixed playlist is about 80, the chances of a random prediction hitting the true change point $\Lambda$ would be $1/80$. The displacement error $\Delta$ of this approach should be larger. However, when we were generating the change points using playlists, we sampled $\Lambda$ from a uniform distribution. Therefore, the results of RP were better than a completely random guess. HMCD-S2 and HMCD-S10 both outperformed the baseline methods.


Table~\ref{tab:rec_results} shows the performance of different recommendation methods. Among the baseline methods, as expected PopRank is the worst performing method in both datasets. However, MC works surprisingly well on Spotify but not so well on LastFM. The main reason could be that Spotify contains public playlists potentially shared among many users with similar transition patterns among albums. LastFM data, on the other hand, contains individual listening logs corresponding to unique actual user activity sequences. In general, results on LastFM are more reliable because it contains actual user data. The performance of BPR is also noticeable compared to sequential recommendation methods such as FOSSIL and GRU. We think that the good performance of BPR is due to its ability to deal with binary or implicit feedback data (a pair-wise approach for ranking). Also, the poor performance of GRU is probably because of long lengths of the sequences of user interactions in both datasets. Usually, GRU works well on session-based data where the information about users' preferences are condensed into sessions with small number of items. Finally, our method in Spotify dataset using SMF-S2 outperforms all of the other methods except MC, and in LastFM, SMF-S3 is the best performing sequential recommendation model.    


\section{Conclusion}

In this work, we introduced the concept of change point detection in order to model the dynamics of user preferences in sequential recommendation tasks. We devised a recommendation framework based on Hidden Markov Models to detect changes in user preferences and use the identified change points to appropriately target generated recommendations. Specifically we extended the standard Viterbi algorithm to detect the change points in sequences of user-item interactions. Using two datasets in music domain, we demonstrated the effectiveness of our methods in terms of both accuracy of change point detection and accuracy of resulting recommendations.



\bibliographystyle{aaai}
\bibliography{paper.bib}

\end{document}